\documentstyle[preprint,aps]{revtex}
\tighten
\draft
\widetext
\input epsf
\preprint{HUTP-99/A019, NUB 3199, EFI-99-11}
\bigskip
\bigskip
\begin{document}
\title{Higher Loop Effects on Unification via Kaluza-Klein Thresholds}
\medskip

\author{Zurab Kakushadze$^{1,2,3}$\footnote{E-mail: 
zurab@string.harvard.edu} and 
Tomasz R. Taylor$^2$\footnote{E-mail: taylor@neu.edu}}

\bigskip
\address{$^1$Jefferson Laboratory of Physics, Harvard University,
Cambridge,  MA 02138\\
$^2$Department of Physics, Northeastern University, Boston, MA 02115\\
$^3$Enrico Fermi Institute, University of Chicago, Chicago, IL 60637}

\date{May 18, 1999}
\bigskip
\medskip
\maketitle

\begin{abstract}
{}We discuss higher loop corrections to gauge coupling renormalization in the
context of gauge coupling unification via Kaluza-Klein thresholds. We show that
in the case of ${\cal N}=1$ supersymmetric compactifications the one-loop 
threshold contributions are dominant, while the higher loop correction are
subleading. This is due to the fact that at heavy Kaluza-Klein levels the
spectrum as well as the interactions are ${\cal N}=2$ supersymmetric. In 
particular, we give two different arguments leading to this result - one is
field theoretic, while the second one utilizes the power of string perturbation
techniques. To illustrate our discussions we perform explicit two-loop 
computations of various corrections to gauge couplings within this framework. 
We also remark on
phenomenological applications of our discussions in the context of TeV-scale
brane world.
\end{abstract}
\pacs{}

\section{Introduction}

{}D-branes \cite{polchi} are likely to play an important role in describing 
nature. In particular, the Standard Model gauge and matter fields may live 
inside of $p\leq 9$ spatial dimensional $p$-branes, while gravity lives in 
a larger (10 or 11) dimensional bulk of the space-time. This ``Brane World'' 
picture\footnote{For recent developments, see, {\em e.g.}, 
\cite{witt,lyk,TeV,dienes,3gen,anto,ST,3gen1,TeVphen,BW,zura}. Large radius
compactifications were originally discussed in \cite{ANT}.} 
{\em a priori} appears
to be a viable scenario, and, based on considerations of gauge and 
gravitational coupling unification, dilaton stabilization and weakness 
of the Standard Model gauge couplings, in \cite{BW} it was actually argued 
to be a likely description of nature. In particular, these phenomenological 
constraints seem to be embeddable in the brane world scenario 
(with the Standard Model fields living on branes with $3<p<9$), which therefore
might provide a coherent picture for describing our universe 
\cite{BW}\footnote{The brane 
world picture in the effective field theory context 
was discussed in \cite{early,shif}.}. 
This is largely due to a much higher degree of flexibility of the brane 
world scenario compared with, say, the old perturbative heterotic framework. 

{}As an example consider the gauge and gravitational couplings 
which in string theory are expected to unify 
(up to an order one factor due to various thresholds \cite{kap,bachas}) 
at the string scale
$M_s=1/\sqrt{\alpha^\prime}$. 
In the brane world scenario {\em a priori} 
the string scale can be anywhere between the electroweak 
scale $M_{ew}$ and the Planck scale $M_P=1/\sqrt{G_N}$ (where $G_N$ is
the Newton's constant). If we assume that the bulk is ten dimensional,  
then the four dimensional gauge and gravitational couplings 
scale as\footnote{For illustrative purposes here we are using 
the corresponding tree-level relations in Type I (or Type I$^\prime$) theory.} 
$\alpha\sim g_s/V_{p-3} M_s^{p-3}$ respectively 
$G_N\sim g_s^2/V_{p-3} V_{9-p} M_s^8$, 
where $g_s$ is the string coupling, and $V_{p-3}$ and $V_{9-p}$ are the 
compactification volumes inside and transverse to 
the $p$-branes, respectively. 
For $3<p<9$ there are two {\em a priori} independent volume factors, 
and, for the fixed gauge coupling $\alpha$ (at the
unification, that is, string scale) and four 
dimensional Planck scale $M_P$, the string scale is
not determined. This observation was used in 
\cite{witt} to argue that the gauge and gravitational 
coupling unification 
problem\footnote{For a review of the gauge and 
gravitational coupling unification problem in 
the perturbative heterotic string context, see, {\em e.g.}, \cite{Dienes}, 
and references therein. 
In the Type I context the discussions on this issue 
can be found in \cite{CKM,BW}.} can be ameliorated in this context 
by lowering the string scale $M_s$ down to the GUT scale 
$M_{\small{GUT}}\approx 2\times 10^{16}~{\mbox{GeV}}$ 
\cite{gut}\footnote{By the GUT scale here we mean the usual 
scale of gauge coupling unification in the MSSM 
obtained by extrapolating the LEP data in the assumption of 
the standard ``desert'' scenario.}. In \cite{lyk} it was 
noticed that $M_s$ can be further lowered all the way down to TeV.

{}In fact, in the brane world picture {\em a priori} the string 
scale can be as low as desired as long as 
it does not directly contradict current experimental data. 
In \cite{TeV} it was proposed that 
$M_s$ as well as\footnote{Note that the string 
scale $M_s$ cannot be too much lower than the fundamental Planck 
scale or else the string coupling $g_s$ as well as 
all the gauge couplings would come out too small 
contradicting the experimental data.} 
the fundamental (10 or 11 dimensional) Planck scale 
can be around TeV. The observed weakness of the four 
dimensional gravitational coupling then requires the presence 
of at least two large ($\gg 1/M_s$) compact directions 
(which are transverse to the branes on which the Standard Model 
fields are localized). A general discussion of possible 
brane world embeddings
of such a scenario was given in \cite{anto,ST,BW}. 
In \cite{TeVphen} various non-trivial phenomenological 
issues were discussed in the context of the TeV-scale 
brane world scenario, and it was argued that this possibility 
does not appear to be automatically ruled out\footnote{TeV-scale 
compactifications were studied in \cite{quiros} in
the context of supersymmetry breaking.}. 

{}However, in such a scenario, as well as in any scenario where $M_s\ll
M_{\small{GUT}}$, the gauge coupling unification at $M_s$ would 
have to arise in a way
drastically different from the usual MSSM unification which
occurs with a remarkable precision \cite{gut}. 
It is then desirable to have a mechanism in the TeV-scale brane 
world scenario for lowering the unification scale. 
Moreover, it would also be necessary to find a concrete 
extension of the MSSM (where this new mechanism is realized) 
such that the unification prediction is just as 
precise as in the MSSM (at least at one loop). 
In fact, one could also require that such an extension {\em explain} why
couplings unify in the MSSM at all, that is, why the 
unification in the MSSM is {\em not} just an ``accident'' 
assuming that the TeV-scale brane world scenario has the pretense of 
replacing the old framework.
 
{}In the brane world picture there
appears to exist a mechanism \cite{dienes} for lowering 
the unification scale. Thus, let the ``size'' $R$ of 
the compact dimensions inside of the $p$-brane 
(where $p>3$) be somewhat large compared with $1/M_s$. 
Then the evolution of the gauge couplings above the Kaluza-Klein (KK) 
threshold $1/R$ is no longer logarithmic but power-like \cite{TV}. 
This observation was used in \cite{dienes} to argue that the gauge 
coupling unification might occur at a scale (which in the brane 
world context would be identified with the string scale) 
much lower than $M_{\small{GUT}}$. 

{}In \cite{zura} a TeV-scale Supersymmetric Standard Model (TSSM) 
was proposed. 
The gauge coupling unification in the TSSM indeed occurs via 
such a higher dimensional mechanism. Moreover, 
the unification in the TSSM is as precise (at one loop) 
as in the MSSM, and occurs in the TeV 
range\footnote{By the TeV range we do {\em not} necessarily 
mean that $M_s\sim 1~{\mbox{TeV}}$. In fact, as was argued in \cite{zura}, 
the gauge coupling unification constraints seem to imply that $M_s$ 
cannot really be lower than $10-100~{\mbox{TeV}}$.}. 
The TSSM also explains why the unification in the MSSM 
is not an accident - if the TSSM is indeed (a part of) the 
correct description of nature above the electroweak scale, 
then the gauge coupling unification in the MSSM is explained 
by the current lack of data which leads to the standard ``desert'' 
assumption. Moreover, as was pointed out in \cite{zura}, after a rather 
systematic search the TSSM is the only (simple) solution for 
the constraints guaranteeing that the gauge couplings unify as precisely 
(at one loop) as in the MSSM.

{}An important question that arises in the context of unification via 
KK thresholds as well as its concrete realization via the TSSM is the
issue of higher loop corrections. The point here is that even though the
unified gauge coupling typically is small, the loop expansion parameter is 
still of order 1 for it is amplified by a large number of heavy KK modes 
running in the loops \cite{zura}. 
In the most general case it is therefore far from being 
obvious that the higher loop corrections are subleading. In fact, if
one considers a generic KK compactification of a higher dimensional theory 
{\em without} supersymmetry, the higher loop corrections are as large as the
one-loop threshold contribution, and therefore the very idea of gauge 
coupling unification via Kaluza-Klein thresholds in this context has no 
predictive power.  

{}However, as was pointed out in \cite{zura}, in the context of 
${\cal N}=1$ supersymmetric 
KK compactifications the situation is drastically different. In this paper 
we elaborate the 
discussions in \cite{zura} as well as \cite{TV}
on the issue of higher loop corrections. In 
particular, we show that the higher loop corrections are subleading compared
with the leading one-loop threshold contribution. Here we give two different
ways of arriving at this result. The first argument is purely field  theoretic
and is along the lines of that given in \cite{zura}, albeit the argument we 
give in this paper is somewhat simpler. The key observation underlying this 
argument is that the heavy KK modes (in certain orbifold compactifications
we consider in this paper) have extended, namely, ${\cal N}=2$ supersymmetry. 
This statement does not only concern the heavy KK spectrum, but also the
{\em interactions} involving only heavy KK modes - the three- and four-point
couplings of heavy KK states are ${\cal N}=2$ supersymmetric. This then 
implies that the leading (in powers of the relevant number of the KK modes) 
contribution at a given higher loop order 
vanishes due to certain ${\cal N}=2$ supersymmetric 
cancellations (recall that the gauge coupling is not renormalized beyond two 
loops in ${\cal N}=2$ supersymmetric theories). The second argument we give 
in this paper is string theoretic and utilizes the power of string perturbation
techniques. In fact, the string theoretic discussion we give in this paper 
allows one to arrive at a clear geometric interpretation of the above 
mentioned cancellations at higher loops via the D-brane picture as well as
string world-sheet expansion in terms of Riemann surfaces of various 
topologies. In this sense, this argument is very much in the spirit of 
(but not exactly the same as) that employed in 
\cite{BKV} to prove finiteness of certain large $N$ gauge theories.

{}To illustrate the formal arguments mentioned above, we explicitly compute 
various two-loop corrections to the gauge couplings
within the above framework. In fact, the cancellations
of heavy KK contributions can be seen explicitly in these two-loop 
computations. We also discuss the general setup for performing higher loop 
computations in the KK theories of this type.

{}The rest of this paper is organized as follows. In section II we 
describe the setup for discussing higher loop corrections 
to the gauge coupling unification via Kaluza-Klein thresholds. In particular, 
we discuss certain orbifold compactifications of higher dimensional theories 
in this context. In section III we discuss one-loop unification 
via Kaluza-Klein thresholds. In particular, we give the general expression for 
the one-loop KK thresholds in the context of the orbifold theories discussed in
section II. In section IV we give the field and string theoretic arguments
which show that the underlying ${\cal N}=2$ supersymmetry 
(at the heavy KK levels) 
indeed implies 
cancellation of the leading higher loop contributions. 
We also discuss the 
general setup for computing the subleading higher loop corrections.
In section
V we give explicit two loop computations in the above context.  
In section VI we remark on various important issues
in the context of unification via Kaluza-Klein thresholds.

\section{Setup}

{}In this section we describe the setup for discussing
higher loop corrections to the gauge coupling unification via Kaluza-Klein
thresholds. In particular, we will discuss certain orbifold
compactifications of higher dimensional theories in this context. For
definiteness, we will focus on compactifications of six dimensional gauge
theories. Generalization to other cases is completely straightforward. 

{}Thus, consider a six dimensional gauge theory living in the world-volume
of some set of coincident D5-branes\footnote{Here we concentrate on
D-brane theories, albeit our discussion is unmodified if we have some other
type of branes.}. Let the space transverse to the D5-branes be a K3
surface. Then we have ${\cal N}=1$ supersymmetric gauge theory living in
the world volume of the D5-branes. In the following discussion, for the most
part, the directions transverse to the branes are not 
important. However, we will still keep track of them to have a clear
geometric interpretation of the four dimensional gauge theory. 
Next, compactify two of the directions
inside of the D5-branes on a two-torus $T^2$. The low energy effective
field theory is then given by the corresponding four dimensional 
${\cal N}=2$ supersymmetric gauge theory. Here by low energies we mean
energies below the compactification scale. Thus, let $T^2=S^1\otimes S^1$,
where both circles have the same radius $R$. Then the compactification
scale is given by the mass scale of the first Kaluza-Klein threshold
$1/R$. Below this scale we have an effective four dimensional gauge theory.

{}We would now like to obtain an ${\cal N}=1$ supersymmetric gauge theory
from this setup. To do this, we will mod out the above theory by a
discrete orbifold group $\Gamma$. Here $\Gamma$ must act on both $T^2$ and
K3 such that $(T^2\otimes{\mbox{K3}})/\Gamma$ is a Calabi-Yau three-fold with
$SU(3)$ holonomy. Note that the action of $\Gamma$ on $T^2$ must be
crystallographic (or else it is not a symmetry of the
theory). This restricts the allowed choices of $\Gamma$ to Abelian cyclic
groups ${\bf Z}_2,{\bf Z}_3,{\bf Z}_4,{\bf Z}_6$. In the following we will
generally refer to $\Gamma$ as ${\bf Z}_M$ with the understanding that $M$
is restricted to the above values.   

{}To discuss the spectrum of the above orbifold model, let us introduce
some notations. First, let $G$ be the gauge group of the original six
dimensional gauge theory. Note that $G$ can be a product group. Let the
matter content of the six dimensional theory be given by hypermultiplets in
various representations of $G$. We will collectively denote these matter
hypermultiplets by ${\widetilde \Phi}$. Each hypermultiplet can be written
as ${\widetilde \Phi}=\Phi\oplus \Phi^\prime$, where $\Phi$ denotes a
chiral multiplet (in the four dimensional sense) with left chirality,
whereas $\Phi^\prime$ denotes the corresponding anti-chiral multiplet
with right chirality. Similarly, let ${\widetilde V}$ be the gauge vector
supermultiplet (of the four dimensional ${\cal N}=2$ gauge theory)
transforming in the adjoint of $G$. It can be written as ${\widetilde
V}=V\oplus \chi$, where $V$ denotes the corresponding ${\cal N}=1$ gauge 
vector supermultiplet (which contains a Weyl fermion with left chirality),
whereas $\chi$ denotes the complement of $V$ in ${\widetilde V}$, that is,
the ${\cal N}=1$ anti-chiral supermultiplet (which contains a Weyl fermion
with right chirality) transforming in the adjoint of $G$.

{}Next, let us discuss the action of the orbifold group $\Gamma\approx{\bf
Z}_M$ on various degrees of freedom. First, we will assume that the action
of $\Gamma$ on the gauge quantum numbers is trivial (that is we choose the
corresponding gauge bundle to be trivial). Thus, orbifolding does {\em not}
break the original gauge group $G$. 

{}Now, let us discuss the breaking of the ${\cal N}=2$
supersymmetry to ${\cal N}=1$ by the orbifold group $\Gamma$. 
Let $g$ be the generator of $\Gamma$. Then the action of $g$ on
the R-parity quantum numbers of states in ${\widetilde V}$ and ${\widetilde
\Phi}$ is the following: $g|V\rangle=|V\rangle$,
$g|\chi\rangle=\omega|\chi\rangle$, $g|\Phi\rangle=|\Phi\rangle$, 
$g|\Phi^\prime\rangle=\omega^{-1}|\Phi^\prime\rangle$, where $\omega\equiv
\exp(2\pi i/M)$. Geometrically this can be understood from the fact that
(up to an obvious choice of convention) $g$ acts on the complex coordinate
$z_1$ on $T^2$ as $gdz_1=\omega dz_1$, while its action on the holomorphic
two-form $\Omega_2\equiv dz_2\wedge dz_3$ on K3 is given by
$g\Omega_2=\omega^{-1}\Omega_2$. Here $z_2,z_3$ are the complex coordinates
on K3. Note that the holomorphic three-form $\Omega_3\equiv dz_1\wedge
dz_2\wedge dz_3$ is invariant under the action of $\Gamma$, which is nothing
but the condition that the quotient $(T^2\otimes{\mbox{K3}})/\Gamma$ is a
Calabi-Yau three-fold.

{}Finally, we turn to the action of $\Gamma$ on the Kaluza-Klein modes
corresponding to the compactification on $T^2$. Let $\gamma^{ij}$, 
$i,j=1,2$, be
the inverse metric\footnote{Thus, in the case of a square torus 
$T^2=S^1\otimes S^1$ the inverse metric reads: $\gamma^{11}=1/R_1^2$, 
$\gamma^{22}=1/R_2^2$, and $\gamma^{12}=\gamma^{21}=0$, 
where $R_1$ and $R_2$ are the radii 
of the two circles.} 
on $T^2$. Then the masses of the KK modes are given by:
\begin{equation}
 M_{\bf m}^2=\gamma^{ij} m_i m_j~.
\end{equation}        
Here ${\bf m}\equiv(m_1,m_2)$, and $m_1,m_2$ are the integer KK momenta
corresponding to the $a$-cycle respectively $b$-cycle on $T^2$. The action
of the generator $g$ of $\Gamma$ on the KK momenta is given by the
corresponding $2\pi/M$ rotation: $g|{\bf m}\rangle=|\theta{\bf
m}\rangle$. Here $\theta$ is the corresponding rotation matrix:
$\theta:m_i\rightarrow {\theta_i}^j m_j$. Note that $\theta^M$ is an identity
matrix. 

{}The zero KK modes with ${\bf m}=(0,0)$ are invariant under the action of
$g$. At heavy KK levels with ${\bf m}\not=(0,0)$ 
we can form linear combinations such that the action
of $g$ is diagonal. These are given by ($\ell=0,1,\dots,M-1$):
\begin{equation} 
 |{\bf m};\ell\rangle\equiv{1\over \sqrt{M}} \sum_{k=0}^{M-1} 
 \omega^{-\ell k}|\theta^k{\bf m}\rangle~.
\end{equation}    
Note that $g|{\bf m};\ell\rangle=\omega^\ell|{\bf m};\ell\rangle$.

{}Now we are ready to determine the spectrum of the orbifold theory. We
must project onto the states invariant under the action of $\Gamma$, that
is, onto the $g$-invariant states. These are given by (${\bf m}\not=(0,0)$)
\begin{eqnarray}
 &&|(0,0)\rangle\otimes |V\rangle~,~~~
  |(0,0)\rangle\otimes |\Phi\rangle~,\nonumber\\
 &&|{\bf m};0\rangle\otimes |V\rangle~,~~~  
  |{\bf m};M-1\rangle\otimes |\chi\rangle~,\nonumber\\
 &&|{\bf m};0\rangle\otimes |\Phi\rangle~,~~~
  |{\bf m};1\rangle\otimes |\Phi^\prime\rangle~.\nonumber
\end{eqnarray}
Here, to avoid overcounting, independent choices of ${\bf m}$ are
restricted to the appropriate conjugacy classes each containing $M$
elements such that they are all related by the corresponding 
$\theta$ rotations.

{}Note that the massless modes are ${\cal N}=1$ supersymmetric. However,
the massive KK modes form ${\cal N}=2$ supermultiplets. Thus, for each
choice of ${\bf m}$ (in the corresponding conjugacy class),  
$(|{\bf m};0\rangle\otimes |V\rangle)\oplus  
(|{\bf m};M-1\rangle\otimes |\chi\rangle)$ forms a (short) 
massive ${\cal N}=2$
supersymmetric vector multiplet. Similarly,  
$(|{\bf m};0\rangle\otimes |\Phi\rangle)\oplus
(|{\bf m};1\rangle\otimes |\Phi^\prime\rangle)$ forms a massive ${\cal
N}=2$ hypermultiplet. In fact, this statement is not just about the massive
KK spectrum, but also holds for massive KK {\em interactions} as well. More
concretely, it is not difficult to see that in the orbifold theory
the three- and four-point 
couplings involving only heavy KK modes are ${\cal N}=2$ supersymmetric
with strengths rescaled by $1/\sqrt{M}$ respectively $1/M$ compared with the
corresponding couplings in the parent ${\cal N}=2$ theory. As we will see in
the next section, this fact has important implications for higher loop
corrections to the gauge coupling renormalization. In particular, it
implies that the higher loop diagrams with only heavy KK modes running in
the loops vanish due to ${\cal N}=2$ supersymmetry. Thus, for a higher loop
correction to be non-vanishing, it is required that it involves at least
one massless line inside of the loops. As we have already mentioned,
massless states do {\em not} have ${\cal N}=2$ supersymmetry but are only
${\cal N}=1$ supersymmetric so that the corresponding higher loop
contributions are (generically) non-vanishing. Nonetheless, as we point
out in the next section, such higher loop diagrams are subleading compared
with the one-loop gauge coupling renormalization due to the KK
thresholds. This is the key reason why the gauge coupling unification via
KK thresholds is meaningful in the present setup.

{}Before we end this section, we would like to point out that we can
introduce additional massless matter 
fields in the above picture such that they do 
not have heavy KK counterparts. We will collectively denote these fields by
$\phi$. Thus, the $\phi$ fields can be localized at the fixed points of the
orbifold. That is, they are truly $3+1$ dimensional in contrast, say, to the
fields ${\widetilde V}$ and ${\widetilde \Phi}$ which can be viewed as
propagating in $5+1$ dimensions (two of which are compact). The fields
$\phi$ then only have ${\cal N}=1$ supersymmetry, and their couplings
to other fields including heavy KK modes are also ${\cal N}=1$
supersymmetric. One of the reasons for introducing such localized matter 
fields is that their presence is typically desirable in concrete
phenomenological model building. 

\section{Unification}
          
{}In this section we discuss one-loop threshold corrections to gauge coupling 
renormalization due to the heavy KK modes. In particular, we give the general
expression for the KK thresholds in the context of the orbifold theories
discussed in the previous section. We then apply these results to one-loop 
gauge coupling unification via Kaluza-Klein thresholds.

{}To begin with, let us clarify what we mean by gauge coupling unification. 
Suppose $G$ is a product gauge group: $G=\bigotimes_a G_a$, where $G_a$ are 
simple subgroups (which can be Abelian or non-Abelian). At the tree level the 
gauge couplings $\alpha_a$ for all subgroups are assumed to be the same: 
$\alpha_a\equiv\alpha$. In string theory this is the case if the entire gauge
group comes from the same set of coincident 
branes. In fact, the tree-level ``unified'' 
gauge coupling $\alpha$ is given by (here we are using the conventions of 
\cite{TASI}):
\begin{equation}
 \alpha=g_s/2v_{p-3}~.
\end{equation} 
Here $g_s$ is the (ten dimensional) string coupling, and $v_{p-3}$ measures
the size of the $p-3$ compact spatial directions inside of the D$p$-branes:
\begin{equation}  
 v_{p-3} \equiv V_{p-3}(M_s/2\pi)^{p-3}\equiv (RM_s)^{p-3}~,
\end{equation}
where $V_{p-3}$ is the actual volume of these $p-3$ compact directions, and 
$M_s=1/\sqrt{\alpha^\prime}$ is the string scale. Note that here $R$ is 
understood only as an ``effective size'' 
of compactification, and in general it 
need not coincide with the actual linear dimension(s) of the compactification
space.  

{}The above tree-level relations are subject to radiative corrections. Here
we focus on one-loop corrections to the gauge couplings 
$\alpha_a$. In particular, we are interested in gauge couplings at low 
energies $\mu\ll 1/R <M_s$. These gauge couplings $\alpha_a(\mu)$ depend upon
the energy scale $\mu$. More concretely, the energy scale dependent one-loop 
corrections to the gauge couplings come from the corresponding infra-red (IR)
divergences due to the massless modes propagating in the loop. Then $\mu$ 
plays the role of the IR cut-off. This gives precisely the familiar field 
theoretic logarithmic evolution of gauge couplings. There are also energy 
scale {\em independent} corrections due to various {\em thresholds}
corresponding to massive modes (with masses $\gg\mu$) propagating in the loop. 
These thresholds come from the heavy KK modes as well as string states such as 
string oscillator modes. At one loop we have:
\begin{equation}
 \alpha_a^{-1}(\mu) =\alpha^{-1} +{b_a\over 2\pi}\ln\left({M_s\over\mu}
 \right)+\Delta_a~,
\end{equation}
where the one-loop $\beta$-function coefficients $b_a$ 
correspond to the massless 
${\cal N}=1$ supersymmetric modes: $b_a=b_a(V)+b_a(\Phi)+b_a(\phi)$ (see
section II for notations).
Here we have chosen the ultra-violet (UV) 
cut-off to be the string scale $M_s$. 
This is equivalent to identifying the gauge coupling unification scale with the
string scale. In particular, if we start from the low energy gauge 
couplings, we can run them up to the string scale by ``integrating in'' the 
heavy KK modes as we go to higher and higher energies. Then, with the above 
convention, the gauge couplings unify at the string scale $M_s$ subject to the 
appropriate choice of the subtraction scheme. In the above approach, 
as we will see in a moment, the choice of the subtraction scheme affects the 
threshold corrections $\Delta_a$.

{}In the following we are interested in the regime where
$(RM_s)^{p-3}\gg 1$. In this case the threshold corrections due to the KK 
modes are large and dominate the string thresholds which generically are of 
order 1 (or smaller) \cite{kap}. We will therefore focus on the KK thresholds.

{}As we will see in a moment, the KK threshold computation in the 
${\cal N}=1$ supersymmetric orbifold theory reduces to that in the parent 
${\cal N}=2$ gauge theory. Let us therefore first consider the ${\cal N}=2$
gauge theory with the superfields ${\widetilde V}$ and ${\widetilde \Phi}$ 
which arises upon compactifying the corresponding six dimensional ${\cal N}=1$ 
supersymmetric gauge theory (living in the world-volume of D5-branes) 
on $T^2$. In the following, for the sake of notational convenience, we will be
a bit more general and treat it as a compactification of a $p+1$ dimensional
theory (living in the world-volume of D$p$-branes) on a 
$(p-3)$-torus $T^{p-3}$. 
Note that $p$ here can take values $p=4,5$, and in the former case we have a 
compactification of a five dimensional gauge theory on a circle (here the 
corresponding orbifold group can only be ${\bf Z}_2$).       
In these notations the KK spectrum of the ${\cal N}=2$ theory consists of 
states with masses $M_{\bf m}^2=\gamma^{ij}m_i m_j$ 
(${\bf m}\equiv(m_1,\dots,m_{p-3})$, $m_i\in{\bf Z}$) with quantum numbers 
${\widetilde V}$ and ${\widetilde \Phi}$. All of the states including the 
massless ones are ${\cal N}=2$ supersymmetric, and only contribute to the 
renormalization of the low energy gauge couplings at one loop - perturbatively
there are no corrections to the gauge couplings beyond one loop, which is due 
to ${\cal N}=2$ supersymmetry. Moreover, in the D-brane context {\em no other}
states contribute to the gauge coupling renormalization. This follows from 
the fact that in perturbative open string theory only BPS states can 
renormalize gauge couplings \cite{lerche,bachas}. In six dimensional 
${\cal N}=1$ open string theories the only BPS states are the massless modes, 
while all the other states are non-BPS as they come in six dimensional 
${\cal N}=2$ (that is, four dimensional ${\cal N}=4$) supermultiplets. 
The latter, however, do not renormalize gauge couplings (in the four 
dimensional gauge theory arising upon compactification on $T^2$). Therefore, 
we are left only with the KK threshold contributions, which we are going to 
discuss next.

{}The computation of the gauge coupling renormalization in this ${\cal N}=2$ 
gauge theory can be performed entirely within the field theory approach. 
In particular, we will treat this theory as a gauge theory with a UV 
cut-off $\Lambda$ containing massless modes plus massive KK states. 
The role of the cut-off $\Lambda$ is to restrict the heavy KK modes to a 
finite subset which essentially amounts to discarding those with masses larger
than $\Lambda$. More precisely, one can employ the standard 
Coleman-Weinberg prescription which gives the following simple result 
\cite{TV}:
\begin{equation}
 \alpha_a^{-1}(\mu)=\alpha^{-1}_a(\Lambda) +{{\widetilde b}_a\over 4\pi}
 \int_{(\xi\Lambda)^{-2}}^{(\xi\mu)^{-2}} {dt\over t}\sum_{{\bf m}}
 \exp(-\pi tM_{\bf m}^2)~.
\end{equation}
Here $\mu$ and $\Lambda$ are the IR respectively UV cut-offs, and we have 
parametrized the subtraction scheme dependence by $\xi$. The one-loop 
$\beta$-function coefficients ${\widetilde b}_a$ are those of the ${\cal N}=2$
theory: ${\widetilde b}_a=b_a({\widetilde V})+b_a({\widetilde \Phi})$ (see 
section II for notations). Next, we will identify $\Lambda$ with the string 
scale $M_s$, and the gauge couplings $\alpha_a(\Lambda)$ with the ``unified''
gauge coupling $\alpha$. Note that this is the only place where string theory
becomes relevant - as we will discuss in the next section, for the above 
prescription to be meaningful, we must assume that above the cut-off the 
theory is finite, which is precisely what we expect above the string scale 
where string theory description takes over. With these identifications, we 
will obtain the expected logarithmic evolution of the low energy 
($\mu\ll 1/R <M_s$) gauge couplings $\alpha_a(\mu)$. This logarithmic 
contribution comes from the massless modes with ${\bf m}=0$:
\begin{equation}
 \alpha^{-1}_a(\mu) =\alpha^{-1} +{{\widetilde b}_a
 \over 2\pi} \ln\left({M_s\over\mu}\right)+
 {\widetilde \Delta}_a~.
\end{equation}   
The IR {\em finite} threshold corrections ${\widetilde \Delta}_a$ are due to 
the massive KK modes with ${\bf m}\not=(0,\dots,0)$. The leading contribution 
(in the regime $(RM_s)^{p-3}\gg 1$) to ${\widetilde \Delta}_a$ can be readily 
evaluated using the Poisson resummation, and the result is given by \cite{TV}:
\begin{equation}
 {\widetilde \Delta}_a={{\widetilde b}_a\over 2\pi}{\xi^{p-3}\over{p-3}}
 (RM_s)^{p-3} - {{\widetilde b}_a\over 2\pi} \ln\left(RM_s\right)+{\cal O}(1)~.
\end{equation}     
Here for the sake of simplicity in the case of $T^2$ we assume that the 
metric $\gamma_{ij}$ 
on $T^2$ satisfies $\gamma_{11}\sim \gamma_{22}\sim \det^{1/2}(\gamma_{ij})=
R^2$ (that is, the complex structure on $T^2$ is ``of order 1'').
Note that the subtraction scheme dependent parameter $\xi$ cannot be 
determined within these considerations alone\footnote{In 
\cite{dienes} the choice of the subtraction scheme was such that
$\xi^{p-3}=\pi^{(p-3)/2}/\Gamma((p-1)/2)$.}. However, in a given theory $\xi$
affects the unification scale $M_s$ (for given values of the low energy gauge 
couplings), and it could, therefore, be determined by fixing $M_s$ via some 
other low energy quantity (such as the Newton's constant $G_N$).

{}Now we are ready to determine the threshold corrections $\Delta_a$ 
in the ${\cal N}=1$ orbifold theory. In fact, we simply have       
\begin{equation}\label{thres}
 \Delta_a={\widetilde \Delta}_a/M~.
\end{equation}
This follows from the fact that the number of the massive KK modes in the 
${\cal N}=1$ theory is $M$ times smaller than in the parent ${\cal N}=2$ 
theory, which is due to the ${\bf Z}_M$ orbifold projection. Note that, as 
we have already mentioned, in (\ref{thres}) we are ignoring other 
${\cal O}(1)$ threshold corrections, namely, those due to heavy string modes.

\section{Higher Loop Corrections}

{}In the previous section we discussed one-loop renormalization of 
gauge couplings due to KK thresholds. In this section we address the issue of 
higher loop corrections. In particular, {\em a priori} it might seem that 
higher loops would destroy the one-loop prediction for the low energy gauge
couplings obtained from the assumption that they unify at the string scale, 
or, equivalently, one-loop unification of gauge couplings via 
Kaluza-Klein thresholds. If so, then the entire framework would have no 
predictive power.

{}There are various ways of thinking about higher loop effects, some of them 
being more precise than others. Thus, one might naively argue that the 
theories discussed in section II are higher dimensional and therefore 
non-renormalizable. This would imply that one has no control over higher loop 
corrections. Fortunately, however, this naive argument is too naive. The key
observation here is that the regimes we are dealing with are such that
the theories under consideration never become higher dimensional. To make this 
statement more precise, let us recall the tree-level 
relation between the four dimensional gauge coupling $\alpha$, the string
coupling $g_s$, the string scale $M_s$ (which is nothing but the UV cut-off
in this context), and the compactification ``size'' $R$:
\begin{equation}
 \alpha=g_s/2(RM_s)^{p-3}~.
\end{equation} 
Note that we are working in the regime where the lowest KK threshold is below
the string scale. In fact, it is much below the string scale: $(RM_s)^{p-3}
\gg 1$. If we take $R$ to be smaller than $1/M_s$, then the 
description in terms
of D$p$-branes becomes inadequate - we have to ``T-dualize'' into a more 
adequate description in terms of lower dimensional branes. Thus, if all $p-3$ 
extra spatial dimensions become smaller than $1/M_s$, the adequate description
is in terms of D3-branes. Now, if we take $R$ to be large compared with 
$1/M_s$, naively this might seem to be sufficient to go into the 
decompactification regime where the theory essentially becomes $p+1$ 
dimensional. However, we wish to keep the four dimensional gauge coupling 
$\alpha$ fixed, so for fixed $g_s$ we would have to decrease $M_s$ while
increasing $R$ such that $RM_s(\gg 1)$ is fixed. This implies that
in the decompactification limit $R\rightarrow \infty$ the UV cut-off 
$\Lambda=M_s$ of the theory goes to zero. That is, the theory never becomes a 
higher dimensional theory (which would be non-renormalizable) 
with a finite cut-off. Another way of phrasing this statement is that we are
dealing with four dimensional gauge theories with a large but {\em finite} 
number $N$ of heavy KK modes. This number $N$ is determined by the cut-off 
{\em vs.} the first KK threshold ratio. That is, we can define $N$ to be 
given by $N=(RM_s)^{p-3}$. Note that here we are implicitly making the 
assumption that above the UV cut-off scale $M_s$ the theory is actually 
finite or else truncation of the KK modes to a finite subset could not be 
justified. This assumption, however, holds in the string theory context, so as
long as we bear in mind that these theories are meaningful only if we embed 
them in a larger theory such as string theory, the above naive arguments about
these theories being higher dimensional and non-renormalizable do not hold.

{}{\em A priori} there is, however, a more serious worry which can be made 
precise. Thus, consider a generic KK compactification (which, in particular,
need not be supersymmetric). Naively it might seem that as long as the 
``unified'' gauge coupling $\alpha$ is small the higher loop corrections are
negligible. This is, however, not the case - the true loop expansion parameter
is {\em not} 
$\alpha/2\pi$ (which would be the case if the effective field theory 
description in terms of just the light modes were adequate all the way up to 
the string scale). Rather, the correct expansion parameter is related to the
string coupling $g_s$. This is due to the fact that although each KK mode 
(including the light modes) couples with the strength of order $\alpha$, there
are many, namely, $N$ KK modes propagating inside of the loops. The true loop
expansion parameter is therefore\footnote{This definition is consistent with 
the fact that in the closed string sector the loop expansion parameter
is $(g_s/4\pi)^2=\lambda^2$, while in the open string sector it is $n_D 
(g_s/4\pi)=n_D \lambda$. Here $n_D$ is the number of D$p$-branes. In the 
gauge theory discussion it arises through the corresponding $\beta$-function 
coefficients. In fact, the statement that $n_D\lambda$ is the loop expansion 
parameter in the open string sector is essentially precise. A more precise 
statement is that the $L$-th loop is weighted by $b_{L-1}\lambda^L$ just
as is the case in the gauge theory language.} 
\begin{equation}
\lambda\equiv \alpha N/2\pi~. 
\end{equation}
This is analogous to 
considering the effective 't Hooft's coupling in large $N$ gauge theories 
\cite{thooft}. As pointed out in \cite{zura}, in concrete phenomenological 
applications the effective coupling $\lambda$ can be of order 1. If so, then
generically one would expect that higher loop threshold corrections are just
as large as one-loop thresholds, that is, that this framework lacks any 
predictive power.

{}However, in \cite{zura} it was pointed out that supersymmetry saves 
the day. That is, in the case of non-supersymmetric theories the above 
argument indeed shows that one has no control over higher loop effects. In 
the supersymmetric case, however, there are subtle cancellations at higher
loops such that one loop thresholds are always dominant. Here we would like
to reiterate the field theoretic 
argument of \cite{zura}, and give another way of arriving at 
the same result using the power of string perturbation techniques.
In the next section we will give explicit two-loop computations which 
illustrate the general arguments presented in the remainder of this section.

\subsection{Field Theory Argument}    

{}In this subsection we first review (a simplified version of) 
the argument of \cite{zura} which shows that
higher loop corrections to the gauge coupling unification via KK thresholds
are indeed subleading compared with the leading one-loop contribution. In the 
following for the sake of simplicity we will omit the subscript $a$ labelling 
the corresponding gauge subgroups $G_a$. On general grounds we expect that the
renormalized gauge coupling at scales $\mu\ll 1/R$ is given by
\begin{equation}
 \alpha^{-1}(\mu)=\alpha^{-1} +f(\mu) +\Delta~,
\end{equation} 
where $\Delta$ is the ($\mu$-independent) contribution due to heavy KK 
thresholds, whereas $f(\mu)$ corresponds to the gauge coupling running. Here we
are interested in estimating the sizes of both $f(\mu)$ and $\Delta$. The 
$L$-loop order contribution $\alpha^{-1}_L (\mu)$ to $\alpha^{-1}(\mu)$ can be
schematically written as
\begin{equation}\label{L-loop}
 \alpha^{-1}_L (\mu)=\alpha^{-1} \sum_{m=0}^L c_{m,L} N^m \left({\alpha\over 
 2\pi}\right)^L~,
\end{equation}
where $m$ counts the number of the loop propagators corresponding to heavy KK
modes with {\em independent} KK momentum summations. (Note that the KK momentum
must be conserved inside of the loop diagram as the two external gauge boson 
lines carry zero KK momentum.) The enhancement factor $N^m$ arises due to $N$ 
heavy KK states propagating in each of the $m$ internal lines. The 
coefficients $c_{m,L}$ generically depend on the energy scale $\mu$ via the 
appropriate IR cut-off. Note that for $N\gg 1$ the coefficients 
$c_{m,L}\sim 1$ (or smaller) for energy scales not too much smaller than 
$1/R$. (Thus, here and in the following we will treat logarithms of the type 
$\ln(RM_s)$ as being of order 1.) 

{}For $\lambda\sim 1$, naively one might expect a large (that is, of order 
$\alpha^{-1}$) contribution to 
$\alpha^{-1}(\mu)$ coming from the term $m=L$ in 
(\ref{L-loop}). 
(Note that this term is $\mu$-independent. More precisely, it is finite in 
the limit $\mu\rightarrow 0$. This is because all of the internal propagators 
in this case correspond to heavy KK states.) However, for $L\geq 2$ this 
contribution actually vanishes due to ${\cal N}=2$
supersymmetry of heavy KK modes. Indeed, as we have already pointed out in 
section II, the spectrum of heavy KK modes is ${\cal N}=2$ supersymmetric. 
Moreover, their interactions (that is, those involving only heavy KK modes
but no massless states) are also ${\cal N}=2$ supersymmetric. This then 
implies that all diagrams with two external gauge bosons involving only heavy
KK modes inside of the loops vanish for gauge couplings are not renormalized
beyond one loop in ${\cal N}=2$ theories. Thus, non-vanishing contributions
can only come from diagrams involving at least one massless internal 
propagator (with massless modes corresponding to $V,\Phi,\phi$). These 
diagrams, however, are suppressed by additional powers of $N$. Thus, we have  
\begin{eqnarray}
 &&\alpha_0^{-1}=\alpha^{-1}\sim N~,\\
 &&\alpha_1^{-1}(\mu)\sim N~,\\
 &&\alpha_{L>1}^{-1}(\mu)\sim \lambda^{L-1}\sim 1~,
\end{eqnarray}
where the estimates here should be understood symbolically (that is, 
we are suppressing the $\mu$ dependence in 
the corresponding contributions). 
This implies that higher loop contributions to the gauge coupling 
renormalization are subleading compared with the one-loop 
contribution, although they are of order 1 so we have to worry 
about the perturbative ``convergence'' issues. 
However, as was argued in \cite{zura} using properties of the holomorphic 
gauge coupling in ${\cal N}=1$ supersymmetric theories, we expect the
perturbation series to converge for $\lambda<\eta_c$, where
$\eta_c\sim 1$ is a model dependent convergence radius (which, in particular, 
depends on the superpotential).  The resummed higher loop corrections 
are then still expected to
be of order 1 and therefore subleading compared with the one-loop 
contribution \cite{zura}. 
 
{}In the cases where $\lambda$ is somewhat smaller than 1 the perturbation 
theory can be trusted. In particular, in various phenomenologically 
applications one might wish to compute higher loop corrections to the gauge 
coupling unification. In section V we will setup the general framework for 
computing two-loop corrections and present various explicit calculations at
the two-loop order. Here, however, we would like to point out a set of simple
rules which seem to be useful in computing higher loop corrections in 
${\cal N}=1$ orbifold theories using the corresponding computations in 
the parent ${\cal N}=2$ theories.

{}It is convenient to normalize gauge couplings so that they are the same
in both the ${\cal N}=1$ and ${\cal N}=2$ theories. At the same time we will
take the $T^{p-3}$ in the case of the ${\cal N}=2$ theory to be 
the same as that 
appearing in the orbifold $T^{p-3}/\Gamma$ in the ${\cal N}=1$ theory. Then 
we have ${\mbox{Vol}}(T^{p-3}/{\bf Z}_M)=V_{p-3}/M$, where $V_{p-3}$ is the 
volume of $T^{p-3}$. To have the low energy gauge couplings the same in the two
theories, we then must take ${\widetilde g}_s=Mg_s$, where ${\widetilde g}_s$ 
and $g_s$ are the string couplings in the ${\cal N}=2$ respectively 
${\cal N}=1$ theories.

{}Having fixed the relative normalization for the gauge couplings in the two 
theories, let us now discuss three- and four-point couplings involving various
states. As we have already pointed out in section II, 
the three-point couplings 
involving only heavy KK states are $1/{\sqrt M}$ times weaker in the 
${\cal N}=1$ theory compared with those in the parent ${\cal N}=2$ theory.
(It then follows that the corresponding factor for the four-point couplings 
involving only heavy KK states is $1/M$.) 
As to the three-point couplings involving some massless states with $V,\Phi$
quantum numbers, they are not affected by the orbifold projection. That is, 
if there is at least one massless state in a three-point coupling,
then this coupling is the same as in the parent ${\cal N}=2$ theory.
On the other hand, the four-point couplings involving one massless state are
reduced by $1/\sqrt{M}$, while the four-point couplings involving two massless
states are unaffected.
Using these rules we can obtain some useful 
information about the relation between 
various diagrams in the two theories. Thus,
consider an $L$-loop diagram with $m>0$ internal propagators corresponding to 
heavy KK modes with independent KK momentum summations. 
For definiteness let us focus on diagrams involving only three-point 
functions. (The discussion straightforwardly generalizes to diagrams involving
four-point functions as well.) It is not 
difficult to see that the value of this diagram in the orbifold theory is
reduced by the factor $(1/M)^{2m-1}$ compared with the same diagram in the 
parent ${\cal N}=2$ theory. This suppression factor arises as follows. 
Note that the heavy KK propagators contribute the suppression factor
$(1/M)^m$ compared with the corresponding diagram in the parent ${\cal N}=2$ 
theory for the number of the heavy KK modes in the orbifold theory is 
$M$ times smaller than in the ${\cal N}=2$ theory. 
On the other hand, there are $2(m-1)$ three-point functions involving
only heavy KK modes, and each of these three-point functions contributes the
suppression factor of $1/\sqrt{M}$. Putting all of this together, we obtain the
above total suppression factor of $(1/M)^{2m-1}$. Note that internal lines
corresponding to massless states are not modified compared with the parent 
${\cal N}=2$ theory except for the fact that some of the original ${\cal N}=2$
massless states are absent in the ${\cal N}=1$ theory due to the orbifold 
projection. This implies that to compute a given diagram with, say, $m>0$, we 
can take the corresponding diagram in the ${\cal N}=2$ theory, discard the 
terms involving massless states projected out by the orbifold action, and 
divide the resulting expression by $M^{2m-1}$. This will give us the answer 
for the corresponding ${\cal N}=1$ computation. In the case of diagrams 
involving only massless internal propagators (that is, for $m=0$), there is 
no suppression factor (unlike in the $m>0$ case), 
and to obtain the answer starting 
from the corresponding ${\cal N}=2$ computation, we only need to discard terms
involving states projected out by the orbifold action. 

The above observations are very useful for 
simplifying higher loop computations in the orbifold theory once the
corresponding computations have been done in the parent ${\cal N}=2$ theory.
This will become clear in the next section where we present some
explicit two-loop computations. We will see that
these simplifications are quite substantial since certain diagrams 
(or, more precisely, combinations thereof) vanish {\em even} if they contain 
massless internal propagators. 
On the other hand, the diagrams involving ``twisted'' states 
cannot be obtained by a simple
orbifold reduction of the ${\cal N}=2$ theory, therefore they must be 
computed on a model-to-model basis.

\subsection{String Theory Argument}

{}In this subsection we give a simple string theory argument which 
allows one to arrive at the above result by utilizing 
the power of string perturbation techniques. Let us start from the parent 
${\cal N}=2$ supersymmetric theory. In string theory language the perturbative
expansion is in terms of Riemann surfaces with various topologies. For the 
sake of simplicity let us focus on oriented Riemann surfaces with $b$ 
boundaries and $g$ handles. In particular, for the sake of simplicity 
we will ignore diagrams with 
cross-caps which are straightforward to incorporate in the following discussion
(see the next subsection for some relevant comments).
Each boundary corresponds to D-branes, whereas handles 
correspond to closed string loops. Note that each diagram is weighted by 
$g_s^{2g-2+b}$, where $g_s$ is the string coupling.

{}Let us first consider the one-loop open 
string diagram with two boundaries and no handles. This is the annulus 
amplitude. In the open string loop channel it is weighted by $g_s^0 N$. The 
enhancement by a factor of $N$ is due to heavy KK modes propagating in the
loop. Alternatively, we can view this diagram in the closed string tree 
channel where a closed string is exchanged between two D-branes whose boundary
states we will denote by $|{\widetilde D}\rangle_0$. 
Note that in the closed string tree 
channel there is no enhancement due to exchange of some large number of 
states - 
all such states are at least as heavy as $\sim M_s$ (the winding modes are 
even heavier with masses $\sim RM_s^2\gg M_s$). This implies that the 
enhancement of the annulus amplitude by a factor of $N$ in the closed string 
tree channel is due to the normalization of D-brane boundary states:
\begin{equation}   
 |{\widetilde D}\rangle_0\sim N^{1/2} |{\widetilde D}^\prime\rangle_0~.
\end{equation}
Here the boundary state $|{\widetilde D}^\prime\rangle_0$ is normalized so that
$_0\langle {\widetilde D}^\prime |{\widetilde D}^\prime\rangle_0\sim 1$.
Note that this normalization factor can be derived directly by computing the
annulus amplitude in the open string loop channel, and then performing the
open-closed duality transformation (that is, the transformation $t\rightarrow
1/t$, with $t$ the proper time on the annulus). Indeed, in the open string loop
channel we have the KK momentum sum which in the closed string tree
channel produces the volume factor $V_{p-3}$ upon the appropriate Poisson 
resummation (recall that $N\sim V_{p-3} M_s^{p-3}$).

{}Let us now consider a diagram with $b$ boundaries and no handles. 
Note that this diagram corresponds to $L=b-1$ loops in the open string 
channel. In the 
closed string tree channel it can be viewed as $b$ D-branes connected by
a closed string tree with $b-2$ three-point vertices corresponding to closed
string interactions. Each of these vertices is weighted by the factor
$g_s/(M_s^{p-3}V_{p-3})^{1/2}\sim g_s/N^{1/2}$, 
where $V_{p-3}$ is the volume of the $p-3$ compact 
directions inside of D$p$-branes. Here we are ignoring the volume of 
transverse directions (that is, of K3, or, more precisely, of K3$/\Gamma$, 
which we assume to be of order 1 in the string units). Thus, such a diagram
is weighted by 
\begin{equation}
 N^{b/2} (g_s/N^{1/2})^{b-2}\sim 
 \alpha^{L-1} N^L~.  
\end{equation}
Here on the l.h.s. the first factor $N^{b/2}$ 
corresponds to $b$ D-branes (or, more precisely,
their boundary states $|{\widetilde D}\rangle_0$ scaling as 
$\sim N^{1/2}$), whereas
the second factor $(g_s/N^{1/2})^{b-2}$ 
comes from the closed string interaction vertices. On the 
r.h.s. we have taken into account that $g_s/N\sim
\alpha$, and $L=b-1$. Thus, this contribution corresponds to the term $m=L$
in the expression (\ref{L-loop}) applied to the ${\cal N}=2$ theory. 

{}Diagrams with handles correspond to subleading terms with $m<L$ in the
language of (\ref{L-loop}). Indeed, each
handle is accompanied by two three-point interaction vertices leading to the 
suppression by $g_s^2/N\sim \alpha^2 N$. That is, diagrams
with $b$ boundaries and $g$ handles are weighted by  
\begin{equation}
 N^{b/2} (g_s/N^{1/2})^{b-2+2g}\sim
 \alpha^{L-1} N^{L-g}~,
\end{equation}
where we have taken into account that $L=b+2g-1$. Thus, we see that $m=L-g$
in the language of (\ref{L-loop}). 
Note that the coefficients $c_{m,L}$ in the
${\cal N}=2$ gauge theory vanish for $L>1$. That is, the only non-vanishing
diagram in this case is that with $b=2,g=0$.

{}Next, let us consider the orbifold theory with ${\cal N}=1$ supersymmetry.  
Here we have $M$ different types of D-branes whose corresponding boundary
states we will denote by $|D\rangle_k$, $k=0,1,\dots,M-1$. The 
{\em untwisted} boundary states $|D\rangle_0$ are coherent states made of
(left-right symmetric) untwisted sector closed string states. They are related
to the boundary states in the parent ${\cal N}=2$ theory via
\begin{equation}
 |D\rangle_0={1\over \sqrt{M}}|{\widetilde D}\rangle_0~.
\end{equation} 
This can be, for instance, seen by noting that the annulus amplitude in the
orbifold theory is obtained from that in the parent theory by inserting
the projection operator (here $g$ is the generator of ${\bf Z}_M$)
\begin{equation}
 {1\over M}\sum_{k=0}^{M-1} g^k
\end{equation}
into the trace over the Hilbert space of open string states (in the loop
channel). The contribution with $k=0$ then corresponds to the closed string 
exchange between the untwisted boundary states in the closed string tree 
channel. This is, however, the same as in the parent theory up to the overall
factor of $1/M$. 

{}As to the contributions corresponding to $k\not=0$, these
in the closed string tree channel map to closed string exchanges between
the corresponding {\em twisted} boundary states $|D\rangle_k$, $k\not=0$. 
The latter are coherent states made of twisted sector closed string states
localized at the {\em fixed points} of the orbifold. Note that unlike the 
untwisted boundary states, the twisted boundary states are normalized so that
there is no $N^{1/2}$ enhancement factor. That is,
\begin{equation}
 _0\langle D|D\rangle_0\sim N~,
\end{equation} 
while for the twisted boundary states ($k\not=0$) we have
\begin{equation}
 _k\langle D|D\rangle_k\sim 1~.
\end{equation}
This can be seen as follows. First, since the closed string states making up
the twisted boundary states are localized at fixed points, these boundary 
states
cannot possibly know about the volume of the compact space - they are fixed
by the local properties of the orbifold in the vicinity of the fixed points.
In particular, even if we take the volume of the orbifold to infinity 
(that is, if we take $V_{p-3}\rightarrow\infty$), the twisted boundary states 
remain the same. Equivalently, we can arrive at the same conclusion by 
noticing that in the open string loop channel a non-trivial twist $g^k$ leaves
only the origin of the KK momentum lattice invariant. This implies that the
corresponding character does {\em not} 
contain a sum over the KK momentum states.
In particular, the open-closed duality transformation (which amounts to taking
$t\rightarrow 1/t$, where $t$ is the proper time on the annulus) does {\em 
not} produce any volume factors (upon the corresponding Poisson resummation). 
Note that the untwisted characters do contain the KK momentum sum in the open
string loop channel, so in the closed string tree channel we get the volume
factor $V_{p-3}$ (which eventually gives rise to the $N^{1/2}$ 
normalization of the untwisted boundary states).

{}To summarize, we have untwisted boundary states with the enhanced 
normalization factor, and also the twisted boundary states without such an 
enhancement factor. The former are the same as in the ${\cal N}=2$ theory
(up to the overall factor of $1/\sqrt{M}$).
The latter are those that carry the information about supersymmetry breaking.

{}Now we are ready to discuss higher loop corrections in the ${\cal N}=1$
theory. Let us start with diagrams without handles. Out of $b$ boundaries
we can have $b_U$ untwisted boundaries plus $b_T$ twisted boundaries. Note 
that if $b_T=0$, then for $b>2$ (that is, at two or higher loops) the 
corresponding contribution to the gauge coupling renormalization vanishes. 
Indeed, the corresponding diagram is the same as in the parent ${\cal N}=2$
supersymmetric theory up to an overall factor given by the appropriate power 
of $M$. 

{}Thus, non-vanishing diagrams for $b>2$ are those with at least one
twisted boundary state. Note that the closed string states propagating along 
the $b_T$ closed string tubes connecting these
boundary states to the interior of the closed string tree are the 
corresponding twisted closed string states. Three-point couplings between
twisted closed string states are weighted by $g_s$ with no additional volume
suppression factors\footnote{More precisely, 
this is the case for twisted states
localized at the same fixed point. Couplings of states localized at different
fixed points are exponentially suppressed with the volume. Such couplings,
however, are not relevant for our discussion here.}. Three-point couplings of 
two ($g^k$ and $g^{M-k}$) twisted sector states with untwisted sector states 
are weighted by $g_s/N^{1/2}$ (just as in the case of three untwisted sector 
states). It is then not difficult to see that the diagrams involving twisted
boundary states can at most be weighted by $g_s^{b-2}\sim \alpha^{L-1} 
N^{L-1}$. For instance, a diagram with $b_U=0$ with {\em all} 
(including those in the interior of the diagram) 
closed string tubes corresponding to twisted sector states would have such a 
weight. Other diagrams either have the same weight or are more suppressed (by
additional factors of $1/N$). Here one must take into account that the ``total
twist'' associated with a given diagram must be trivial, that is, if $k_i$,
$i=1,\dots,b$, label twists corresponding to the boundary states, we must
have 
\begin{equation}
\sum_{i=1}^b k_i=0~({\mbox{mod}}~M)~.
\end{equation}
Thus, the diagrams that ``know'' about reduction of supersymmetry from 
${\cal N}=2$ to ${\cal N}=1$ are all subleading (as they are weighted
by $\alpha^{L-1} N^{L-1} \sim \lambda^{L-1}<1$) compared with the leading 
one-loop diagram (with $b=2$) weighted by $N\gg 1$. The reader can easily 
verify that adding handles does not change this conclusion. Thus, we have
arrived at the result that the one-loop gauge coupling renormalization is 
dominant ($\sim N$) compared with the higher loop corrections (which are at 
best of order 1).

{}Here we would like to point out that the above argument is very much in the
spirit of (but not exactly the same as) that employed in \cite{BKV} to prove
finiteness of certain large $N$ gauge theories.

\subsection{Comments}

{}Before we end this section, we would like to make a few clarifying
remarks. First, the string theory argument we gave in the previous
subsection makes use of the by now well appreciated fact that string
perturbation theory is a very efficient way of organizing various gauge
theory diagrams. So string theory in this argument is only a tool - at the
end of the day we must take a limit $M_s\rightarrow\infty$ (more precisely,
$M_s\gg 1/R \gg\mu$), and the heavy string modes do not play any role
except for providing a proper UV cut-off for the resulting gauge theory
computation. 

{}Next, in the argument of the previous subsection we only considered
Riemann surfaces with boundaries and handles but no cross-caps. 
Cross-caps are boundary states corresponding to the orientifold planes. 
In the context of {\em perturbative} compactifications of open plus 
closed string theories the latter are required to 
cancel various tadpoles. Tadpole cancellation implies UV finiteness of the 
theory which, in the field theory language, would result in cancellation of all
power-like threshold corrections to the gauge couplings. 
This is due to a cancellation between the annulus 
(a cylinder with two boundaries) and the
M{\"o}bius strip (a cylinder with one boundary and one cross-cap)
amplitudes \cite{bachas,bach}.
In the closed string tree-channel this corresponds to cancellation of the
tadpoles due to massless states corresponding to the zero winding modes
that couple to D-branes wrapped on a torus.

{}Thus, as was already pointed out in \cite{zura,zura1}, unification via 
Kaluza-Klein thresholds should be considered in the context of 
{\em non-perturbative} orientifolds where perturbative tadpoles do {\em not}
cancel. The finiteness of the theory is then due additional {\em 
non-perturbative}
states (which can 
arise at both massless as well as massive levels) which provide
an effective cut-off in the theory just as in perturbative heterotic 
compactifications where the modular integration is restricted to the 
fundamental domain which excludes extreme UV momenta (that is, those 
above a certain energy scale $\sim M_s$). In fact, 
(some) non-perturbative
orientifolds, examples of which have been recently discussed in \cite{NPO}, 
can be viewed as hybrid compactifications sharing the features of both 
perturbative heterotic and orientifold compactifications albeit such 
compactifications are non-perturbative from both heterotic and orientifold 
viewpoints. As to the gauge
theory computation, in this context
it must be regularized by introducing a UV cut-off which we
identify with the string scale. 
Such regulator (which does not arise in perturbative orientifolds \cite{bach})
is expected to
arise in {\em non-perturbative} orientifolds 
once we take into account non-perturbative (from
the orientifold viewpoint) string states which do not have a perturbative
description in terms of, say, open strings ending on D-branes. These states
are expected from Type I-heterotic duality - thus, already in ten
dimensions there are infinitely many BPS states (charged under
${\mbox{Spin}}(32)/{\bf Z}_2$) on the heterotic side which cannot be seen
perturbatively on the Type I side. 

{}Thus, in the above setup, that is, in the context of non-perturbative 
orientifolds, gauge coupling renormalization can be computed in the
gauge theory context assuming the appropriate cut-off at the string scale
$M_s$ (more precisely, at some scale of order $M_s$ - we have to deal with
the subtraction scheme ambiguities). The reason why this UV cut-off scale
should be around $M_s$ is clear - quantum gravity becomes important at
energies of order of the (fundamental) Planck scale $M_P\sim M_s/g_s$, which
is of order of the string scale for $g_s\sim 1$.

\section{Explicit Two-Loop Computations}

{}In this section we present a computation of two-loop corrections to
$U(1)$ gauge couplings, for the D5-brane orbifold models
described in section II. This Abelian two-loop case is completely sufficient 
to illustrate some general features of higher loop corrections.

{}We will use the standard background field method and
evaluate the corresponding photon vacuum polarization diagrams.
Our starting point are the diagrams of
${\cal N}=1$ supersymmetric QED in $D=6$ dimensions.
The diagrams describing the compactified ${\cal N}=2$ parent
theory contain internal propagators with momenta of the form
$(p,{\bf m}/R)$, where ${\bf m}/R$ is the KK component, a two-dimensional
lattice vector. The $d^4p$ momentum loop
integrals are UV divergent and require a cut-off.
We will apply dimensional regularization
and continue the divergent integrals to $d=4-\varepsilon$ dimensions.
The UV divergences appear then as $1/\varepsilon$ poles; these are subtracted
in the renormalization procedure, leaving a finite result which
depends on the dimensional regularization scale $\Lambda$. The scale $\Lambda$ 
is a UV cut-off which
(in an appropriate subtraction scheme) can be identified with the 
string scale \cite{kap,gt}.
However, we are not interested here in the details of UV pole subtractions,
therefore we consider ``bare'' diagrams without the UV counterterms.

{}The ${\cal N}=2$ supersymmetric two-loop vacuum polarization diagrams (note 
that here $q^2=\mu^2$)
\begin{equation}
 \Pi_{\mu\nu}(q)=(q_{\mu}q_{\nu}-g_{\mu\nu}q^2)\Pi_2(q^2)
\end{equation}
fall into 3 classes,
$F_2$, $S_2$ and $Y_2$, involving respectively:\\
$\bullet$ hypermultiplet fermion loops;\\
$\bullet$ hypermultiplet scalar loops;\\
$\bullet$ SUSY Yukawa couplings of gauginos to hypermultiplets.\\
These diagrams are depicted in Fig.1. Note that 
all traces of gamma matrices and contractions of Lorentz indices are
performed in $D=6$ dimensions, ensuring
the right counting of fermionic degrees of 
freedom as well as 
incorporating the Yukawa couplings of four-dimensional scalar 
photons to hypermultiplet fermions on the same 
footing as other gauge couplings.
As a result of a straightforward 
computation\footnote{We used the symbolic manipulation
program FORM Version 1.1 written by J. Vermaseren.} we obtain the 
corresponding contributions to $\Pi_2(q^2)$
as the integrals over 3 Feynman parameters, $x$, $y$ and $z$:
\begin{eqnarray}
F_2 &=& 0~,\nonumber\\
S_2 &=& \alpha\Gamma (\varepsilon){(4\pi N)^{\varepsilon}
\over(2\pi)^3}
\sum_{\bf m,n}
 \int_0^1\!\!\int_0^1\!\!
 \int_0^1\! dx\,dy\,dz\,
{2z(1-2zy)(1-z)^{-1+\varepsilon/2}[x(1-x)]^{-\varepsilon/2}\over [{(
{\bf m}+{\bf n}x)^2\over
x(1-x)}(1-z)+{\bf n}^2+q^2R^2zy(1-zy)]^{\varepsilon}}~,\label{scalar}\\
Y_2 &=&-S_2~,\nonumber
\end{eqnarray}
where $N=(\Lambda R)^2$.
Note that the diagrams $F_2$ are the same as the diagrams describing
a version of $D=4$ non-supersymmetric QED obtained by a toroidal
compactification of the ``minimal'' $D=6$ QED with one charged ``electron''. 
These diagrams cancel among themselves hence
there is no two-loop contribution to the $\beta$-function in such a theory. 
Furthermore the electron mass is not renormalized at one loop.

{}The sum of all three
contributions,
\begin{equation}
\Pi_2(q^2)=F_2+S_2+Y_2=0~.
\end{equation}
In this way, we find a complete cancellation
of two-loop corrections to the gauge coupling. Of course, the cancellation of
$1/\varepsilon$ poles is a consequence of the known result,
$b_2=0$ in ${\cal N}=2$ SUSY gauge theories. Our two-loop computation
provides further evidence for the {\em complete} cancellation
of higher loop corrections to gauge couplings.

{}We now proceed to the ${\cal N}=1$ supersymmetric orbifold theory.
As argued in the previous section, the leading large $N$ contribution 
[$m=L=2$ in the  notation of (\ref{L-loop})], which is due to
diagrams involving {\em massive\/} KK excitations propagating 
in {\em all\/} internal lines, is absent. This is due to ${\cal N}=2$ 
supersymmetry of the massive spectrum which is responsible for the
cancellations encountered above. The subleading contribution, $m=1$,
arises from the diagrams involving one massless propagator while the 
$m=0$ contribution is due to purely massless diagrams. However, here
${\cal N}=2$ cancellations also 
result in the vanishing of certain contributions.

{}In the case under consideration, the two-loop 
diagrams involving {\em one\/} massless propagator
with the $\Phi$ quantum numbers add up to 
zero. This can be seen as follows. First consider the corresponding 
diagrams in the ${\cal N}=2$ theory where the massless states carry 
the ${\widetilde \Phi}$ quantum numbers. If we ignore the external lines 
(corresponding to gauge bosons), then we have two more internal lines (both 
of which correspond to heavy KK modes) - one with ${\widetilde \Phi}$ quantum 
numbers, and the other one with the ${\widetilde V}$ quantum numbers. It is
convenient to denote such a diagram by ${\cal A}({\widetilde\Phi}_0,
{\widetilde \Phi}_{\bf m},{\widetilde V}_{-{\bf m}})$, where subscripts 
indicate the KK momenta.
In the ${\cal N}=1$ language we have ${\widetilde \Phi}=\Phi\oplus
\Phi^\prime$, and ${\widetilde V}=V\oplus\chi$ (see section II), so that the
above ${\cal N}=2$ diagram can be written as:
\begin{eqnarray}
 {\cal A}({\widetilde\Phi}_0,
 {\widetilde \Phi}_{\bf m},{\widetilde V}_{-{\bf m}})=&&
 {\cal A}(\Phi_0,
 \Phi_{\bf m},V_{-{\bf m}})+
 {\cal A}(\Phi_0^\prime,
 \Phi^\prime_{\bf m},V_{-{\bf m}})+\nonumber\\
 &&~~~{\cal A}(\Phi_0, \Phi^\prime_{\bf m},\chi_{-{\bf m}})+
 {\cal A}(\Phi_0^\prime,
 \Phi_{\bf m}, \chi_{-{\bf m}})~.
\end{eqnarray} 
Note that by the left-right symmetry ($\Phi\leftrightarrow\Phi^\prime$) of the
theory, we have ${\cal A}(\Phi_0,\Phi_{\bf m},V_{-{\bf m}})=
{\cal A}(\Phi_0^\prime, \Phi^\prime_{\bf m},V_{-{\bf m}})$, and 
${\cal A}(\Phi_0, \Phi^\prime_{\bf m},\chi_{-{\bf m}})=
{\cal A}(\Phi_0^\prime,\Phi_{\bf m}, \chi_{-{\bf m}})$. 
On the other hand, the above
${\cal N}=2$ diagram vanishes: ${\cal A}({\widetilde\Phi}_0,
{\widetilde \Phi}_{\bf m},{\widetilde V}_{-{\bf m}})=0$. This implies that
\begin{equation}\label{vanish}
 {\cal A}(\Phi_0,
 \Phi_{\bf m},V_{-{\bf m}})=-
 {\cal A}(\Phi_0, \Phi^\prime_{\bf m},\chi_{-{\bf m}})~.
\end{equation}   

{}Next, let us see what happens to the above diagrams upon the ${\bf Z}_M$ 
orbifold projection (that is, let us discuss the corresponding diagrams in the
context of the ${\cal N}=1$ theory). We must discard the diagrams involving 
$\Phi^\prime_0$ as the latter states are projected out. We therefore conclude
that after orbifolding the diagram ${\cal A}({\widetilde\Phi}_0,
{\widetilde \Phi}_{\bf m},{\widetilde V}_{-{\bf m}})$ reduces to
\begin{equation}
 {\cal B}={1\over M}\left({\cal A}(\Phi_0,
 \Phi_{\bf m},V_{-{\bf m}})+
 {\cal A}(\Phi_0,
 \Phi^\prime_{\bf m},\chi_{-{\bf m}})\right)=0~.
\end{equation}    
Here the factor of $1/M$ arises due to the fact that in the ${\cal N}=1$ theory
we must restrict the summation over the KK modes with KK momenta in the 
corresponding conjugacy classes (see section II). 
\footnote{This is precisely the factor $(1/M)^{2m-1}$ 
discussed in the previous section, specified to the case of $m=1$.} 
On the other hand, the ${\cal N}=1$ diagram ${\cal B}$ still vanishes due to
(\ref{vanish}). Thus, we have shown that the two-loop diagrams with only one
massless internal line involving the states with the $\Phi$ quantum numbers 
indeed add up to zero.

{}The above discussion implies that non-vanishing diagrams with only one 
massless internal line must be such that the corresponding massless states
carry the $V$ quantum numbers. In the following, we will evaluate
these diagrams; they involve one massless internal line with 
the $V$ quantum numbers
plus heavy KK lines with $\Phi,\Phi^\prime$ quantum numbers. The computation
can actually be simplified by relating the corresponding diagrams in the 
${\cal N}=1$ theory to their ${\cal N}=2$ counterparts. Let us therefore
consider these ${\cal N}=2$ diagrams which we will refer to as 
${\cal A}({\widetilde\Phi}_{\bf m},
{\widetilde \Phi}_{-{\bf m}},{\widetilde V}_0)$.
In the ${\cal N}=1$ language this ${\cal N}=2$ diagram can be written as:
\begin{eqnarray}
 {\cal A}({\widetilde\Phi}_{\bf m},
 {\widetilde \Phi}_{-{\bf m}},{\widetilde V}_0)=&&
 {\cal A}(\Phi_{\bf m},
 \Phi_{-{\bf m}},V_0)+
 {\cal A}(\Phi^\prime_{\bf m},
 \Phi^\prime_{-{\bf m}},V_0)+\nonumber\\
 &&~~~{\cal A}(\Phi_{\bf m}, \Phi^\prime_{-{\bf m}},\chi_0)+
 {\cal A}(\Phi^\prime_{\bf m},
 \Phi_{-{\bf m}}, \chi_0)~.
\end{eqnarray} 
Note that by the $\Phi\leftrightarrow\Phi^\prime$ symmetry
we have ${\cal A}(\Phi_{\bf m},\Phi_{-{\bf m}},V_0)=
{\cal A}(\Phi^\prime_{\bf m}, \Phi^\prime_{-{\bf m}},V_0)$, and 
${\cal A}(\Phi_{\bf m}, \Phi^\prime_{-{\bf m}},\chi_0)=
{\cal A}(\Phi^\prime_{\bf m},\Phi_{-{\bf m}}, \chi_0)$. 
On the other hand, the above
${\cal N}=2$ diagram vanishes: ${\cal A}({\widetilde\Phi}_{\bf m},
{\widetilde \Phi}_{-{\bf m}},{\widetilde V}_0)=0$. This implies that
\begin{equation}\label{vanish1}
 {\cal A}(\Phi_{\bf m},
 \Phi_{-{\bf m}},V_0)=-
 {\cal A}(\Phi_{\bf m}, \Phi^\prime_{-{\bf m}},\chi_0)~.
\end{equation}   

{}Let us now extract the corresponding ${\cal N}=1$ diagrams, which we will 
refer to as ${\cal C}$, surviving the orbifold projection. Since $\chi_0$ is 
projected out by the ${\bf Z}_M$ orbifold, we have 
\begin{equation}
 {\cal C}={1\over M}\left({\cal A}(\Phi_{\bf m},
 \Phi_{-{\bf m}},V_0)+
 {\cal A}(\Phi^\prime_{\bf m},
 \Phi^\prime_{-{\bf m}},V_0)\right)={2\over M}{\cal A}(\Phi_{\bf m},
 \Phi_{-{\bf m}},V_0)~.
\end{equation}
Thus, we see that non-vanishing diagrams of this type are the same as those
in ${\cal N}=1$ supersymmetric QED (with massive charged superfields). Note 
that the corresponding diagram with all massless internal lines is given by 
\begin{equation}
 {\cal C}_0={\cal A}(\Phi_0,\Phi_0,V_0)~.
\end{equation}

{}Also, in practice, we found it most convenient to start from
the original ${\cal N}=2$ diagrams and subtract the contributions involving
$\chi_0$ component of the vector multiplet, {\em i.e.}, the scalar photon
and one of the two photinos. Thus, in particular, we eliminated
the contributions of scalar photons in $S_2$ and $F_2$ by subtracting
the corresponding polarization components of the $D=6$ gauge boson,
and adjusted the overall factor in $Y_2$ to eliminate
the redundant photino propagator. 
First, we write down
the final result for purely massless diagrams, with all internal
lines corresponding to ${\bf m} ={\bf n}={\bf 0}$:
\begin{equation}
\Pi^{\bf 0}(q^2)=\alpha {(\Lambda^2/q^2)^{\varepsilon}\over 2(2\pi)^3}
\bigg[\frac{1}{\varepsilon}
+\ln (4\pi)-\gamma +\frac{7}{2}\bigg]~,
\end{equation}
where $\gamma$ is the Euler's constant. It is easy to check that the
coefficient of the $1/\varepsilon$ pole does correctly reproduce the
two-loop coefficient of ${\cal N}=1$ supersymmetric QED. The diagrams
involving only one massless propagator are given by more complicated
expressions involving Feynman parameter integrals such as in (\ref{scalar}).
However in the most interesting ``low energy, large radius'' limit,
$\Lambda^2/q^2\gg N \gg 1$, these integrals simplify and the result
can be written as a momentum-independent correction:
\begin{equation}
\Pi^{\bf m}=\alpha\frac{2}{M}\sum_{\bf m}{(N/{\bf m}^2)^{\varepsilon}
 \over 2(2\pi)^3}
\bigg[\frac{1}{\varepsilon}
+\ln (4\pi)-\gamma +\frac{1}{2}\bigg]+\delta^{\bf m}~.\label{pip}
\end{equation}
In the above expression we have isolated the contribution
\begin{equation}
\delta^{\bf m}
 =-\alpha
 \frac{4}{M}\sum_{\bf m}{(N/{\bf m}^2)^{\varepsilon}\over 2(2\pi)^3}
\bigg[\frac{1}{\varepsilon}
+\ln (4\pi)-\gamma +\frac{1}{2}\bigg]~,
\end{equation}
which arises from the diagrams involving self-energy corrections to
internal lines, and contains the UV divergence associated with the 
mass renormalization at the one-loop level. The finite part of
$\delta^{\bf m}$ should be combined with the one-loop threshold correction.
It renormalizes the mass scale of the KK excitations, {\em i.e.}, 
the compactification
radius $R$, or, equivalently, the number $N$ of the KK excitations below
the UV cut-off scale. On the other hand, the first term in (\ref{pip})
represents a genuine two-loop threshold correction proportional
to the two-loop $\beta$-function of ${\cal N}=1$ supersymmetric theory
describing ${\cal N}=2$ KK excitations coupled to the ${\cal N}=1$
massless gauge sector.
{}Note that in the above two-loop expression for the heavy KK threshold
contributions to the gauge coupling renormalization the summation over
the KK momenta ${\bf m}$ is cut-off from above by $N=(\Lambda R)^2$. After the
summation the leading two-loop threshold is of order $\alpha N\sim 1$, 
which is subleading compared with the leading one-loop threshold 
correction $\sim N$.

\section{Discussions}

{}In this section we would like to discuss various issues which, we believe,
have important implications in phenomenological applications of gauge coupling
unification via Kaluza-Klein thresholds. Some of these points have been 
previously stressed in \cite{zura}, while some others are new and concern 
various developments subsequent to \cite{zura}. 

{}One of the key points we would like to emphasize is that supersymmetry is
crucial for the arguments of section IV to go through. In particular, without
supersymmetry, namely, without ${\cal N}=2$ supersymmetry at the heavy KK 
levels, the cancellations leading to the fact that higher loop corrections to 
the gauge couplings are suppressed compared with the leading one-loop threshold
contribution would not be possible. That is, in the context of KK 
compactifications of higher dimensional theories without supersymmetry the 
very idea of gauge coupling unification via Kaluza-Klein thresholds would have
no predictive power - higher loop corrections in such a setup would
be as large as one-loop thresholds, hence no control over the former.

{}Even in the supersymmetric case the fact that higher loop corrections are 
subleading compared with the one-loop thresholds is only specific to gauge 
coupling renormalization. Thus, the same does {\em not} hold for higher
point couplings. In particular, the Yukawa couplings receive large higher
loop corrections, and therefore one no longer has predictions for Yukawa
coupling unification such as the $b-\tau$ unification.

{}Since supersymmetry at the heavy KK levels is so crucial, one has to worry
about the effects of supersymmetry breaking which must eventually take place.
Thus, the supersymmetry breaking scale (by which we mean the typical scale of
soft scalar masses) must be low enough compared with the KK threshold scale
$1/R$ or else the corrections due to supersymmetry breaking at the heavy KK 
levels would be too large and spoil the above mentioned cancellations due to
${\cal N}=2$ supersymmetry. This implies that the the string scale $M_s\gg 1/R$
cannot really be brought down to a few TeV in this context. (In the particular
model of \cite{zura} the relation between the string scale $M_s$ and the KK 
threshold $1/R$ is fixed by the unification constraint to be $M_s\sim 40/R$ in
the case of D4-branes, and $M_s\sim 6/R$ in the case of D5-branes.) In fact,
these considerations suggest that the string scale cannot be lower than 
$10-100~{\mbox{TeV}}$.  

{}To implement gauge coupling unification via Kaluza-Klein thresholds in the 
phenomenological context a concrete model is required. Such a model, 
which in \cite{zura1} 
was called TSSM (TeV-scale Supersymmetric Standard Model),
was proposed in \cite{zura}. In fact, it was stressed in \cite{zura} that this 
model 
was the only solution to the unification constraints found in \cite{zura}.
Here we would like to discuss some issues concerning this point.

{}First, in the MSSM the gauge coupling running is given by:
\begin{equation}
 \alpha_a^{-1} (\mu)=\alpha_{GUT} +{b_a^*\over 2\pi} \ln\left({M_{GUT}
 \over \mu}\right)~,
\end{equation}  
where $\alpha_{GUT}\approx 1/24$ is the unified gauge coupling,
$M_{GUT}\approx 2\times 10^{16}~{\mbox{GeV}}$ \cite{gut} is 
the unification scale, and
$b^*_a$, $a=1,2,3$, are the $SU(3)_c\otimes SU(2)_w\otimes U(1)_Y$
$\beta$-function
coefficients ($b^*_1=33/5$, $b^*_2=1$, $b^*_3=-3$, where we have used 
the standard
normalization $\alpha_1=(5/3)\alpha_Y$). In order for the (one-loop) 
unification via the KK thresholds to be just as precise as in the MSSM, it is 
required that the ${\cal N}=2$ $\beta$-function coefficients 
${\widetilde b}_a$ 
at the massive KK levels (see section III) satisfy the following relation:
\begin{equation}\label{nu}
 \nu_{ab}~{\mbox{is independent of $a,b$, where}}~
 \nu_{ab}\equiv{{{\widetilde b}_a
 -{\widetilde b}_b}\over {b^*_a-b^*_b}}~{\mbox {for}}~a\not=b.
\end{equation} 
This condition is satisfied in the TSSM. 
In fact, in the TSSM we have $\nu_{ab}=1~
\forall a\not=b$. As was explained in 
\cite{zura}, the TSSM was the only solution
found there for the unification constraint (\ref{nu}). 
Here we point out that the key assumption
here is that all three gauge subgroups of 
$SU(3)_c\otimes SU(2)_w\otimes U(1)_Y$
come from the {\em same} set of coincident branes. In this case in the string 
theory context we expect all three gauge couplings to be the same at the 
string 
scale (which is identified with the unification scale).

{}Recently, in \cite{carone,quiros} 
a set of models different from the TSSM was 
discussed. According to \cite{carone,quiros}, in these models the (one-loop)
gauge coupling unification is as precise as in the MSSM (and, therefore, TSSM).
Here, however, we would like to point out that there is {\em no} unifications
in the models of \cite{carone,quiros}. 
The point here is that in \cite{carone,quiros}
additional ``solutions'' to the unification constraints 
were found by relaxing the
requirement that all the gauge subgroups come from the same set of coincident 
branes. For instance, {\em a priori} one could imagine that 
$SU(3)_c\otimes U(1)_Y$ subgroups 
are localized on the same set of coincident D$p$-branes
(with $p=4,5$), while $SU(2)_w$ is localized on a fixed point of the orbifold
(such localization, for instance, could be achieved by having other 
branes stuck 
at the orbifold fixed points in the corresponding compact directions). However,
in such cases the gauge couplings of $SU(3)_c\otimes U(1)_Y$ 
on the one hand, and
$SU(2)_w$ on the other hand generically are {\em not} expected to be the same
at the string scale. In 
certain cases equality of these gauge couplings 
can be achieved by some fine tuning, but there is {\em no} 
unification prediction here. In this sense such a scenario is no different from
that proposed in \cite{ST} where various subgroups of the Standard Model gauge 
group come from different sets of branes (and then unification requires fine 
tuning of the corresponding compactification volumes). 
Thus, the TSSM of \cite{zura} (along with its straightforward variations)
is indeed
the only known solution that satisfied the unification constraints.

{}Here another remark is in order. Note that the consistent orbifold reduction
of a higher dimensional gauge theory requires that the heavy KK spectrum is
${\cal N}=2$ supersymmetric. Moreover, suppose we have chiral multiplets $\Phi$
at the massless level. Then at the massive level we must have the corresponding
hypermultiplet (we have been referring to these hypermultiplets as 
${\widetilde \Phi}$) for {\em each} massless chiral multiplet
(with certain degeneracy depending on the order 
$M$ of the orbifold group ${\bf Z}_M$ and the choice of $T^{p-3}$). 
This requirement, which is necessary for the consistency (in particular, 
unitarity)
of the theory at the heavy KK levels is {\em not} met by the models of 
\cite{dienes}. Thus, these models are not completely consistent. Similarly, 
albeit the unification is not a property in 
models of \cite{carone,quiros}, some care is needed
when considering such constructions. In particular, if, say, $SU(2)_w$ is 
localized on a fixed point of the orbifold, 
so must be {\em all} the matter fields
charged under $SU(2)_w$ (which is {\em not} the case in some of the models of 
\cite{carone}) - this is the consequence of the corresponding flux 
conservation requirement.

{}Finally, we would like to comment on the ``UV sensitivity'' issue in the 
context
of gauge coupling unification via Kaluza-Klein thresholds. Thus, naively it 
might 
seem that a small shift in the low energy gauge couplings would spoil the 
unification prediction (whose sensitivity to such shifts is power-like), and 
therefore the gauge coupling unification is too ``UV sensitive'' to be 
predictive.
However, such a viewpoint might be a bit misleading. Thus, consider the TSSM. 
If the unification occurs in the MSSM, then it also occurs in the TSSM, 
and just as
precisely (at one loop) as in the MSSM. Higher loop corrections, 
as we have shown 
in this paper, are small. This implies that such a unification scenario is 
predictive in the context of a concrete model, namely, the TSSM. This is 
essentially due to the fact that the TSSM {\em explains} \cite{zura} 
why the usual logarithmic unification in the MSSM is not an accident 
(assuming that the TSSM is the correct 
description of physics above the electroweak scale), but is rather related 
to the 
lack of experimental data which prompts one to assume the standard ``desert 
scenario'' and extrapolate the gauge couplings all the way to the GUT scale 
$M_{GUT}$.

\acknowledgments

{}We would like to thank Costas Bachas, Gia Dvali and Pran Nath
for valuable discussions. 
This work was supported in part by the grant NSF PHY-96-02074. The work
of Z.K. was also partially supported by the DOE 1994 OJI award. Z.K.
would also like to thank Albert and Ribena Yu for financial support.

\begin{figure}[t]
\epsfxsize=16 cm
\epsfbox{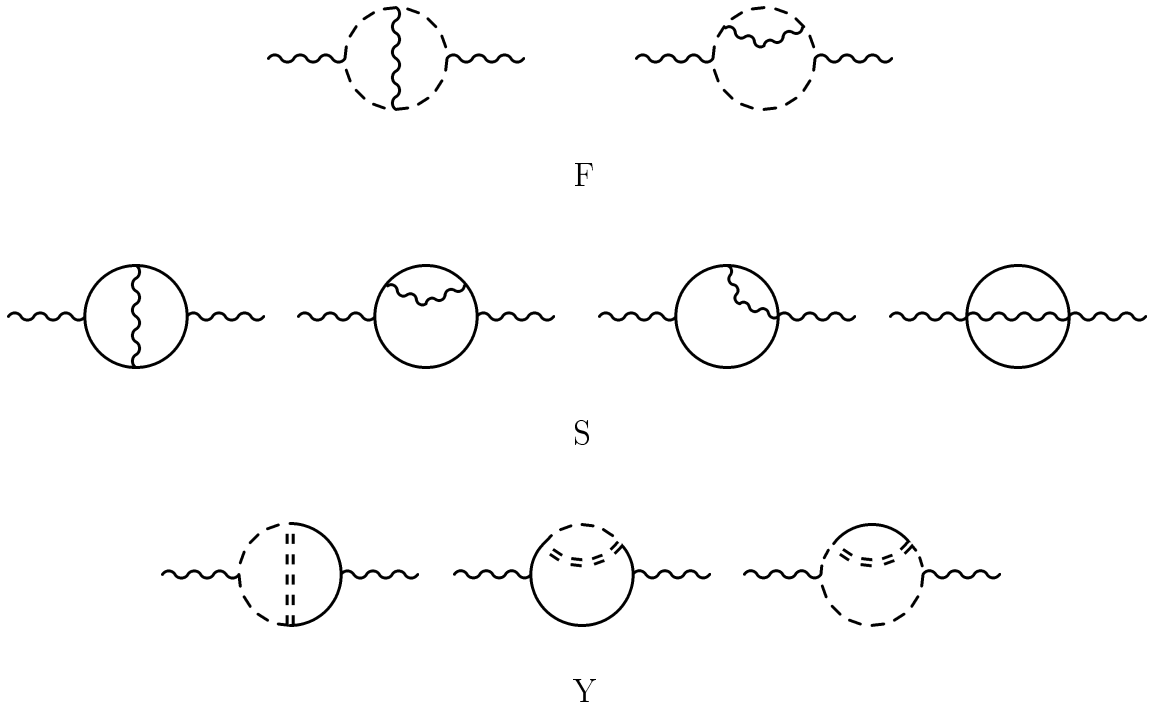}
\bigskip
\caption{Various two-loop gauge theory diagrams discussed in section V. 
Solid lines are hypermultiplet scalars, dashed lines are hypermultiplet 
fermions, double-dashed lines are gauginos, and wavy lines are gauge bosons.
Note that tadpole-like contributions to the scalar propagators are not shown.
Also, one should weight the above diagrams with the appropriate symmetry 
factors.}
\end{figure}

\end{document}